\documentclass{sig-alternate-05-2015}

\PassOptionsToPackage{hyphens}{url}\usepackage[draft]{hyperref} 
\usepackage{graphicx}
\usepackage{listings}
\usepackage{makecell}
\usepackage[latin1]{inputenc}
\usepackage[english]{babel}

\begin{document}

\title{Glassbox: Dynamic Analysis Platform for Malware Android Applications on Real Devices}

\numberofauthors{3}

\author{
\alignauthor
Paul Irolla \titlenote{PhD student}\\
\affaddr{\'{E}cole d'ingénieurs du monde numérique (ESIEA)}\\
\affaddr{Laboratoire de cryptologie et virologie opérationnelles (CVO Lab)}\\
\affaddr{38 rue des Docteurs Calmette et Guérin 53000 Laval, France}\\
\email{irolla[at]esiea[dot]fr}
\alignauthor
Eric Filiol \titlenote{Thesis director}\\
\affaddr{\'{E}cole d'ingénieurs du monde numérique (ESIEA)}\\
\affaddr{Laboratoire de cryptologie et virologie opérationnelles (CVO Lab)}\\
\affaddr{38 rue des Docteurs Calmette et Guérin 53000 Laval, France}\\
\email{filiol[at]esiea[dot]fr}
}

\date{27 July 2016}

\maketitle

\begin{abstract}
\textit{Android} is the most widely used smartphone OS with 82.8\% market share in 2015 \cite{marketshare}. 
It is therefore the most widely targeted system by malware authors. To detect these 
malicious applications before they are installed on users phones, we need an automated analysis. 
Researchers rely on dynamic analysis to extract malware behaviors and often use emulators to do so. 
However, using emulators lead to new issues. Currently emulators cannot emulate SIM card, 
camera and microphone - components that are likely to be used by malware applications. 
Moreover, malware may detect emulation and as a result it does not execute the payload to prevent the analysis. 
Finally, emulation suffers from inherent slowness and causes more application crashes 
than real devices. Dealing with virtual device evasion is a never-ending war and comes with a 
non-negligible computation cost \cite{andrubis}. To overcome this state of affairs, we propose a system 
that does not use virtual devices for analysing malware behavior.

\textbf{\textit{Glassbox}} is a functional prototype for the dynamic analysis of malware applications. 
It executes applications on real devices in a monitored and controlled environment. It is a fully automated 
system that installs, tests and extracts features from the application for further analysis. The environment is 
controlled in a way that \textit{Glassbox} neither suffers from malware nor becomes an infection vector through the
control of web requests, calls and SMS/MMS. The features extracted are Java calls, 
system calls and both encrypted and non encrypted web requests. In this paper, 
we present the architecture of the platform and we compare it with existing \textit{Android} dynamic 
analysis platforms.

Lastly, we evaluate the capacity of \textit{Glassbox} to trigger application behaviors by measuring the 
\textit{average coverage of basic blocks} on the \textbf{\textit{AndroCoverage}} dataset \cite{androcoverage}. 
We show that it executes on average 13.52\% more \textit{basic blocks} than the \textit{Monkey} program.
\end{abstract}


\section{Introduction}
 
Google reacted to the rise of malware with a dynamic analysis platform, named \textit{Bouncer} \cite{bouncer}, 
that analyzes applications before the release on Google Play. This security model is centralized and acts 
before the distribution of applications. Whereas this system suffers from limitations - like virtual device 
evasion - it has helped to reduce the malware invasion about 40\% \cite{bouncer}.

\textit{Android} antivirus companies use another centralized security model which acts after the distribution of applications. 
Because applications have access to restricted resources and permissions, antivirus programs cannot perform their 
analysis - as it often requires root permissions and extensive resources. Hence, the static analysis is externalized  
onto the company servers. As a result, it can give quick responses - each application being 
analyzed just once. This is a shift of security model for the common user toward centralization.   

Users have been used to personal antivirus for their personal computer. This security model does not allow much room for 
manoeuvre because any antivirus needs to be quick enough for not bothering the user - otherwise another quicker antivirus 
will be chosen. Whereas antivirus have implemented heuristic algorithms, they are rather limited by the security 
model. Hence, the security model shifting is an opportunity for building more complex systems that require more resources 
to run. It enables security systems to use advanced research techniques like behavioral detection with dynamic analysis, or 
detection with features similarity from static analysis.

Malware authors made their strategy evolve with the rise of \textit{Bouncer} and other dynamic analysis systems. They have started to hide 
the payload execution with emulation detection or/and the requirement of an user interaction. For example the reverse 
of the sample \cite{evasionsample} shows that malware are using emulation evasion. Emulators settings
can be modified to mimic the appearance of a real device but there are lots of ways of detecting \textit{Android} emulation. 
Actually the tool \textit{Morpheus} \cite{morpheus} proves us this war is already lost as the authors found around 10 000 heuristics to 
detect \textit{Android} emulation. So trying to modify the emulator to look real is probably a waste of time. In such condition, we need to redefine the
problematic.

This is why we are presenting \textit{Glassbox}, a dynamic analysis platform for \textit{Android} malware applications on real devices. 
\textit{Glassbox} is an environment for the controlled execution of applications, where the \textit{Android} OS and the network are monitored 
and have the capacity to block some actions of the analyzed application. This environment is paired with \textit{Smart Monkey}, 
a program that automates the installation, the testing of applications and the cleaning of the environment afterwards. The objective
of \textit{Glassbox} is to collect features for machine learning algorithm, to classify applications as malware or as benign. 
In the following sections we will present the related work about dynamic analysis systems, we will expose the architecture of both \textit{Glassbox}
and \textit{Smart Monkey}. Finally, we will present the \textit{average coverage of basic blocks} of \textit{Smart Monkey} on the \textit{AndroCoverage Dataset} \cite{androcoverage}.

\section{Related Work}

Such dynamic analysis systems started being designed since 2010 \cite{aasandbox} by the academic research, to circumvent the 
limitations of static analysis - namely, code morphism and obfuscation. Since that time, many systems have been released. 
For this study we have built a classification of a part of these systems, presented in Figure \ref{fig:sota} and Figure \ref{fig:sotalegend}.
The classification takes into account three categories : the dynamic features collected by the analysis, the strategies set 
in order to automate application testing and finally the use of real devices in dynamic analysis systems history. 
We discuss the results on the following sub-sections.
 
\begin{figure*}[!ht]\centering
\caption{Comparative state of the art of dynamic analysis systems.}
  \scriptsize
  \begin{center}
    \begin{tabular}{ | c | c | c | c | c | }
      \hline
      \textbf{Reference} & \textbf{Tool name} & \textbf{Dynamic features used} & \textbf{App testing strategies} & \textbf{Objectives \& comments} \\ 
      \hline
      \makecell{Thomas Bläsing \\et al. 2010\\\cite{aasandbox}} & \makecell{AASandbox} & \makecell{System calls (name)} & \makecell{Monkey} & \makecell{Data for malware/benign \\classification\\\# Virtual device}\\ 
      \hline
      \makecell{Iker Burguera \\et al. 2011\\\cite{crowdroid}} & \makecell{Crowdroid} & \makecell{System calls (name)} & \makecell{Crowdsourced app interactions} & \makecell{Data for malware/benign \\classification\\\# Real device}\\ 
      \hline
      \makecell{Cong Zheng \\et al. 2012\\\cite{smartdroid}} & \makecell{SmartDroid} & \makecell{Taint tracking, +?} & \makecell{UI brute force\\Restriction of execution\\to targeted activities} & \makecell{Data for classification \\or manual analysis\\\# Virtual device}\\ 
      \hline
      \makecell{Lok Kwong Yan \\et al. 2012\\\cite{droidscope}} & \makecell{DroidScope} & \makecell{System call (all content)\\Java calls (all content)\\Taint tracking} & \makecell{-} & \makecell{Data for classification \\or manual analysis\\\# Virtual device}\\ 
      \hline
      \makecell{Vaibhav Rastogi \\et al. 2013\\\cite{appsplayground}} & \makecell{AppsPlayground} & \makecell{Taint tracking\\Targeted \textit{Android} API Java calls} & \makecell{Monkey\\UI brute force\\Broadcast events\\Text fields filling} & \makecell{Malware/benign classification\\\# Virtual device}\\ 
      \hline
      \makecell{Martina Lindorfer \\et al. 2014\\\cite{andrubis}} & \makecell{Andrubis} & \makecell{App Java calls (all content)\\System calls (name, +?)\\Shared libraries targeted calls \\(name, +?)\\Taint tracking\\DNS/HTTP/FTP/SMTP/IRC \\(all content)} & \makecell{Monkey\\Broadcast events\\All possible app services\\All possible app activities} & \makecell{Data for classification \\or manual analysis\\\# Virtual device}\\ 
      \hline
      \makecell{Mingyuan Xia \\et al. 2015\\\cite{appaudit}} & \makecell{AppAudit} & \makecell{Taint tracking} & \makecell{$\sim$} & \makecell{Malware/benign classification\\Data leaks detector\\\# Symbolic execution}\\ 
      \hline
      \makecell{Vitor Monte Afonso \\et al. 2014\\\cite{bresiliens}} & \makecell{-} & \makecell{Targeted \textit{Android} API Java calls \\(name)\\System calls (name)} & \makecell{Monkey\\Broadcast events} & \makecell{Malware/benign classification\\96.66\% accuracy\\\# Virtual device}\\ 
      \hline
      \makecell{Kimberly Tam \\et al. 2015\\\cite{copperdroid}} & \makecell{CopperDroid} & \makecell{System calls (all content)\\Binder data} & \makecell{Broadcast events\\Text fields filling, +?} & \makecell{Data for classification \\or manual analysis\\\# Virtual device}\\ 
      \hline
      \makecell{Lifan Xu \\et al. 2016\\\cite{hadm}} & \makecell{HADM} & \makecell{System call (name)} & \makecell{Monkey\\Broadcast events} & \makecell{Malware/benign classification\\87.3\% accuracy\\\# Virtual device}\\ 
      \hline
      \makecell{Marko Dimja{\v{s}}evi{\'c} \\et al. 2016\\\cite{maline}} & \makecell{Maline} & \makecell{System call (name)} & \makecell{Monkey\\Broadcast events} & \makecell{Malware/benign classification\\96\% accuracy\\\# Virtual device}\\ 
      \hline
      \makecell{Michelle Y. Wong \\et al. 2016\\\cite{intellidroid}} & \makecell{IntelliDroid} & \makecell{Taint tracking} & \makecell{Targeted inputs leading to\\suspicious \textit{Android} API calls} & \makecell{Data for classification \\or manual analysis\\\# Virtual device}\\ 
      \hline
      \makecell{-} & \makecell{Glassbox} & \makecell{Java calls (name)\\System calls (name)\\HTTP/HTTPS requests \\(all content)} & \makecell{Monkey\\UI brute force\\Broadcast events\\Real SMS/Call\\Text fields filling} & \makecell{Data for malware/benign \\classification\\\# Real device}\\
      \hline
    \end{tabular}
  \end{center}
  \normalsize
\label{fig:sota}
\end{figure*}

\begin{figure*}[!ht]\centering
\caption{Comparative state of the art of dynamic analysis systems - legend.}
  \scriptsize
  \begin{center}
    \begin{tabular}{ | c | c | }
      \hline
      \makecell{+?} & \makecell{The paper is not clear enough on those details\\and we cannot be sure that it is an exhaustive list}\\
      \hline
      \makecell{call (name)} & \makecell{Only the name of the call is used, in order to get the appearance frequency}\\
      \hline
      \makecell{\#} & \makecell{Comment}\\
      \hline
      \makecell{-} & \makecell{No data}\\
      \hline
      \makecell{$\sim$} & \makecell{Data exists but is irrelevant for this study}\\
      \hline
    \end{tabular}
  \end{center}
  \normalsize
\label{fig:sotalegend}

\scriptsize
\noindent{}Note on the difference between \textit{data for classification} and simply \textit{classification} on the \textit{objectives} column:\\
Many papers, as ours, only present the dynamical analysis system but do not present an analysis of the results on collected features. This can be posponed for another paper or
made by other researchers. For this kind of papers, the objective of the system presented is to produce \textit{data for classification}. For the other ones 
the data is used on a classification algorithm and the results are presented in the paper.  
\normalsize
\end{figure*}

\subsection{Features Analyzed}

Since the rise of \textit{Android} dynamic analysis systems, the use of system calls have been the leading approach. System calls are the functions
of the kernel space, available to the user space. It gives the capacity to manipulate hard drive files or to control processes.
System calls can describe a program behaviors, from a low level perspective. The retrieval of those calls can be achieved in two ways, mainly:

\begin{itemize} 
  \item[\textbullet] \textbf{Virtual Machine Introspection} - This is a technique available for emulators, which enables the host to 
  monitor the guest. It cannot be detected by the guest since it is out of its reach and it is therefore convenient for security analysis.
  \textit{Andrubis} \cite{andrubis}, \textit{CopperDroid} \cite{copperdroid} and \textit{DroidScope} \cite{droidscope} take advantage of VMI to retrieve, unseen by the 
  target malware, all systems calls done by the guest \textit{Android} virtual machine.  
  \item[\textbullet] \textbf{Strace/ptrace} - \textit{Strace} is a Linux utility for debugging processes. It can monitor system calls, signal deliveries
  and changes of process state. \textit{Strace} use the \textit{ptrace} system call to monitor another process memory and registers. This second method is by 
  far the simplest and the most straightforward one as the only task here is the automation of the \textit{strace} execution. Moreover, it targets directly
  the system calls of the application we need to. That is why this method has been adopted in most of the literature, namely \textit{Crowdroid} \cite{crowdroid}, 
  \textit{HADM} \cite{hadm}, \textit{Maline} \cite{maline} and \cite{bresiliens}. We have also chosen to use the \textit{strace} utility for system calls 
  monitoring. Despite the theoretical possibility of a malware to detect that it is being debugged, we found no evidence about this.
\end{itemize}

\noindent
System calls seem to give great results for classification. \textit{Maline} reported 96\% accuracy rate and \textit{HADM} 87.3\% both with syscalls 
frequencies only. Actually syscalls capture low level behaviors of both Java code and native code.

The second most collected feature is taint tracking information as it reveals data leakage. It works by the instrumentation of the \textit{Dalvik VM} interpreter. The information 
we do not want to leak is called a \textit{source}. Some \textit{source} of personal data are tainted, like the phone number or the contacts list. Each time a tainted \textit{source}
 or value is used in a method call, the \textit{DVM} interpreter taints the returned value. With this simple mechanism, we can observe the propagation of the tainted information regardless
 of its transformations. A function that enable to transmit an information outside of the system is called a \textit{sink}, like network requests or SMS. If a tainted value is used in a 
 \textit{sink}, it means data \textit{source} has leaked. It enables to detect data leakage even if this data have been ciphered or encoded. An application that leaks data is not necessarily a malware, 
as data leakage is the business of both malware and user tracking frameworks in commercial applications - which constitutes essentially a large part of "\textit{goodware}" applications. 
Whereas this feature gives useful insights on the application behaviors for manual analysis, its utility for automatic malware detection needs to be proved. Moreover the implementation 
and execution of taint tracking is costly, which leads us not to choose this feature for now.  

Java calls is another feature of interest as it captures an explicit behavior of the application. There are several ways to collect them:

\begin{itemize} 
  \item[\textbullet] \textbf{Application instrumentation} - This strategy does not need any modification of the \textit{Android} source code and is not dependent on \textit{Android} version. The 
  application can be modified in order to dump targeted method parameters and return values. \textit{APImonitor} \cite{apimonitor} is a tool that enables the instrumentation of targeted
  Java calls. It reverses the application into \textit{smali}, a human friendly format equivalent to the Java bytecode, with the \textit{baksmali} \cite{smali} utility. Then, it adds
  monitor routines around the targeted calls and it recompiles the code with the \textit{smali} \cite{smali} utility. This strategy is used by the authors of \cite{bresiliens}. 
   \item[\textbullet] \textbf{DVM/ART instrumentation} - The \textit{DVM} (\textit{Dalvik Virtual Machine}) or \textit{ART} (\textit{Android RunTime}, \textit{Android} API version $\ge$ 4.4)  
  is the system that interprets and executes all the application instructions. All Java calls converge to this component. Hence, by hooking the execution of \textit{DVM/ART}, one can 
  monitor and control all Java calls, their arguments and their return values. That implies the modification of \textit{Android} source code and its compilation to a custom \textit{ROM}. 
  This is the strategy we chose to use for collecting Java calls. We prefer this method for keeping the application behaviors pristine, and particularly not inducing additional bugs.
  \textit{Andrubis} \cite{andrubis} and \textit{DroidScope} \cite{droidscope} use similar approaches for tracing method calls. 
\end{itemize}

For the last features, they are highly marginal. Here, \textit{Andrubis} reported the retrieval of targeted shared library calls. Another data are the network communications.
Only \textit{Andrubis} reported the utilization of features from network communications, but without any further details. Our system makes use of \textit{Panoptes} \cite{autocitation} for 
gathering plain text and encrypted web communications, the process will be described in the \textit{{Network control \& monitoring}} part.

\subsection{Automated Testing Strategies}

Dynamic analysis does not consist of launching the application and waiting the malware to show its malicious behaviors off. Malware are using logic bombs for hiding the payload. 
Logic bombs are a malicious piece of code that is executed after a condition is triggered. It means we need to test each application as a real user could have done it. For achieving this objective, several strategies have been used in the past:

\begin{itemize} 
  \item[\textbullet] \textbf{Black Box testing strategies} - This class of strategies does not take the application source code into account, it focuses on sending inputs in the application without 
  any prior information. This is the commonly used strategy. \textit{Monkey} \cite{monkey} is a dedicated tool created by \textit{Google} for this task. It generates
  random events in a fast pace. Events range from system events (home/wifi/bluetooth/sound volume etc.) to navigation events (motion, click). Because of its capacity to quickly explore applications 
  activities, it has been used by most of dynamic analysis systems (\textit{AASandbox} \cite{aasandbox}, \textit{AppsPlayground} \cite{appsplayground}, \textit{Andrubis} \cite{andrubis}, 
  \cite{bresiliens}, \textit{HADM} \cite{hadm}, \textit{Maline} \cite{maline}). \textit{Monkey} is sometimes confused with \textit{Monkey Runner} \cite{monkeyrunner} in the literature, which 
  is a python library for writing \textit{Android} test routines.
  \item[\textbullet] \textbf{White Box testing strategies} - This class of strategies takes the application source code into account. It focuses on sending specific inputs in the application for 
  triggering targeted code paths. It requires the information from the static analysis of the application. Parsing the code is needed, to find the target methods and all their triggering conditions. 
  \textit{SmartDroid} \cite{smartdroid} and \textit{IntelliDroid} \cite{intellidroid} determine all paths to sensitive API calls, then execute one of the paths to the target with dynamic analysis. 
  Another kind of \textit{White Box} testing strategy is \textit{symbolic execution} where dynamic analysis is done by simulating the execution of the application static code. \textit{AppAudit} 
  \cite{appaudit} uses this technique for finding data leaks with symbolic taint tracking.  
  \item[\textbullet] \textbf{Grey Box testing strategies} - This class of strategies partially takes the applications source code into account. It focuses on testing all visible inputs the application
  declares or displays (UI). It usually takes the output of the application to generate the next inputs. \textit{Andrubis} uses a \textit{Grey Box} strategy when it tests all possible application services
  and activities, because they get the information from the application manifest. \textit{AppsPlayground} also uses \textit{Grey Box} testing with its \textit{Intelligent Execution} where windows, widgets,
  and objects are uniquely identified to know when an object has already been explored. We use a similar strategy in \textit{Glassbox}.     
\end{itemize}

\subsection{Real Devices}

The use of real devices for dynamic analysis started with \textit{Crowdroid} \cite{crowdroid}, a crowdsourced based analysis. Whereas this approach give good results, one cannot ask users to
execute real malware on their personal device. So this system can only be an option, when we have already a trained machine learning algorithm, to find malware in the wild. 

\textit{BareDroid} is a system which manages real devices in large scale for dynamical analysis. Whereas \textit{BareDroid} \cite{baredroid} cannot be considered as a dynamical analysis 
system, because it does not analyse applications, it brought two major results for our study. First, real devices for dynamic analysis systems are a scalable solution financially and
in excution time compared to virtual devices. Second, using real devices drastically improves features detected for malware families that often rely on emulator evasion 
like \textit{Android.HeHe}, \textit{Android Pincer}, and \textit{OBAD}.
 
\section{Architecture overview}

\begin{figure*}[!ht]\centering
\begin{center}
  \includegraphics{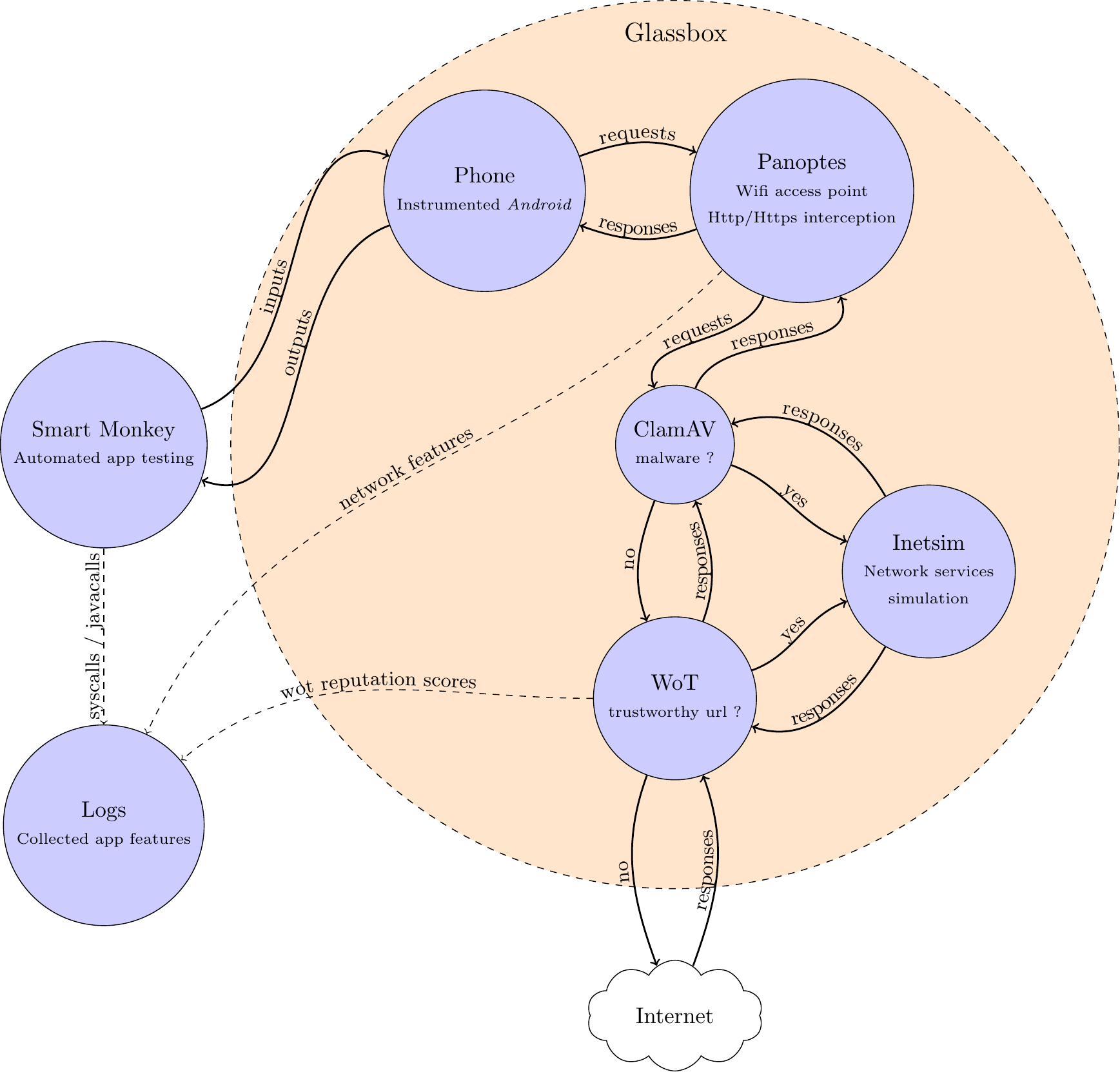}
\end{center}
\caption{Glassbox - Architecture overview.}
\label{fig:overview}
\end{figure*}

\textit{Glassbox} (Figure \ref{fig:overview}) is a modular system distributed among one or several phones and a computer. Each part is detailed in the following sections.

\subsection{Android Instrumentation}

A custom \textit{Android} OS has been made, based on the \textit{\textit{Android} Open Source Project (AOSP)} \cite{aosp}. The objective here is to log dynamically each Java call of 
a targeted application. This involves hooking these calls, at a point where all of them pass through. We instrumented \textit{ART (Android RunTime)} \cite{art}, the \textit{Android} managed runtime 
system that executes application instructions. With the default parameters, we found that \textit{ART} (at least until \textit{Android Marshmallow}) have the following important behaviors 
for our study:

\begin{itemize} 
  \item[\textbullet] The first time \textit{Android} is launched, Java \textit{Android} API libraries and applications are optimized and compiled to a native code format called \textit{OAT} \cite{oat}. 
  \item[\textbullet] Each time a new application is installed, it is optimized and compiled to \textit{OAT} format.
  \item[\textbullet] Java methods can be executed in three ways: by an \textit{OAT} \textit{JUMP} instruction to the method address, by the \textit{ART} interpreter
  for non-compiled methods (debugging purposes mostly), or via the \textit{Binder} for invoking a method from another process or with Java \textit{Reflection}.
  Details on the \textit{Android Binder} can be found in \cite{binder} chapter 4.
\end{itemize}

\noindent{}A straightforward way of hooking Java calls is to instrument the \textit{ART} interpreter. Unfortunately only a few calls are executed through it because most of the code is compiled 
into \textit{OAT} and therefore it is not interpreted. We forced all calls to be interpreted by disabling several optimizations. The first one is the disabling of the compilation to \textit{OAT}. 
That leads calls to be interpreted before executed. But others optimizations mechanisms comes into play, namely \textit{direct branching} and \textit{inlining}.

\begin{figure*}[!ht]\centering
\begin{center}
  \includegraphics[scale=0.7]{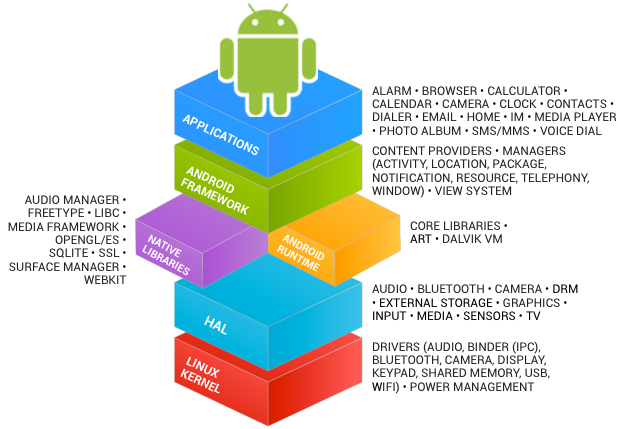}
\end{center}
\caption{Android architecture overview}
\label{fig:Android}
\end{figure*}

The boot classpath contains the \textit{Android} framework (Figure \ref{fig:Android}) and core libraries. They are always compiled in \textit{OAT} resulting in a \textit{boot.oat} file.
This file is mapped into memory by the \textit{Zygote} process, started at the initialisation of \textit{Android}. For launching an application, the main activity is given at \textit{Zygote} in 
parameters. When \textit{Zygote} is called that way, it forks and starts the given activity. It means that any application has access to same instance of the \textit{Android} framework 
and core libraries. \textit{Direct branching} is an optimisation that replaces framework/core method calls by their actual address in memory. So the calls does not pass through the 
interpreter. That optimisation is disabled.

Then \textit{inlining} is an optimisation that replaces short and frequently used methods with their actual code. Although, it slightly increases the application size in memory,
runtime performance are increased. As there is no method any more, it cannot be hooked in the \textit{ART} interpreter. That optimisation is also disabled. 

A monitoring routine is added to \textit{ART} interpreter that logs any method call from a targeted application \textit{pid}. A sample of a capture of Java calls is shown in annexes.
All these modifications overload the global execution of \textit{Android}. Whereas it is not noticeable for most of the applications, on gaming applications are visibly slow down by this approach. 

Lastly the phone is shipped with a real \textit{SIM} card for luring malware payloads with SMS/MMS/Calls. Many malware may use it for stealing money with premium numbers, 
and because we use a real \textit{SIM} card it would actually cost us money. We modified the telephony framework of the \textit{Android} API to reject all outgoing communications 
except for our own phone number. When a forbidden call is made, the calling UI pops and closes after one second around. This way does not crash applications that rely on calls and SMS.     

\subsection{Network Control \& Monitoring}

All communications of the instrumented phone pass through a transparent \textit{SSL/TLS} interception proxy behind a wifi access point. This is set by \textit{Panoptes} \cite{autocitation}. 
To understand how it works we need to describe a part of the \textit{TLS} handshake. Here is the regular behavior of a \textit{https} request on \textit{Android}:

\textit{Android} have a keystore of all root certificates the system trusts. When a \textit{SSL/TLS} request is initialized, the requested server send its certificate. 
It contains identifying informations - like the domain name that must verify the contacted domain name - and a signature that can only be decrypted with the right root CA. 
The server certificate is tested with each trusted root CA, and if one matches the communication is accepted. 
Extended information on the \textit{TLS} handshake can be found with the RFC 2246 memo \cite{tls}. 

For our interception system to work, a \textit{SSL/TLS} root certificate from a custom certification authority (CA) is implanted in the keystore of \textit{Android}.
When the device requests a \textit{https} web page, the request goes through the proxy. It is parsed and a new one is initialised to be sent to the original recipient. 
The response is encapsulated in a new \textit{SSL/TLS} response signed by our custom certificate. This custom certificate is dynamically generated with the recipient identifying 
informations and our custom root CA private key. As the communication is signed by an authority of certification that is known by the client, \textit{Android} accepts it without 
any warning. Finally, all \textit{HTTP/HTTPS} communications are logged and a report can be generated which is convenient for manual analysis if needed.

This system has been extended to support manipulation of requests. The objective is to restrict the proliferation of malware and the damage that it may produce. As \textit{Glassbox}
runs malware, it may have a negative impact on its environment. An extreme mean could be to disconnect the system from internet but we would see less or no malicious behavior at all 
for numerous applications. Our design is a trade-off between safety and behavior detection:    

\begin{itemize} 
  \item[\textbullet] \textit{ClamAV} \cite{clamav} is used to detect known malware sent through network. If a malware is detected, the payload is removed from the request and is 
  redirected to \textit{Inetsim} \cite{inetsim}, a network services simulation server that replies consistently to the requests. It forbids the communication between the application 
  and internet without crashing it. 
  \item[\textbullet] For all other requests, we assess the reputation of the domain name or \textit{IP} address with the \textit{Web of Trust (WoT)} \cite{wot} API. \textit{WoT} 
  is a browser extension that filters urls based on different reputation rating. These rating come mainly from the users. If the request contacts a known address with a good reputation, we 
  forbid the application under test to reach it, then it is redirected to \textit{Inetsim}. The advantages are twofold. The application cannot damage a respectable website, and it 
  pre-filters behaviors for classification.
\end{itemize}

Finally features from communications content are collected and the \textit{WoT} reputation scores as well.
A sample of a network capture is shown in the annexes.

\subsection{Automated Application Testing}

\textit{Smart Monkey} is an automated testing program based on \textit{Grey Box} strategies. The context of the application is determined at runtime for the automatic exploration. 
We use \textit{UIautomator} \cite{uiautomator}, a tool that can dump the hierarchy tree of the current UI elements present on screen. It enable us to monitor variables of 
each UI element at runtime. For a smart exploration, we need to know if we have already processed an element. Unfortunately, elements do not carry such an unique identifier. 
Nonetheless, we found that elements can be identified to some degree:

\begin{itemize} 
  \item[\textbullet] \textbf{Strong identification} - Elements can have an associated \textit{ID string} that developers set. Concatenated to the current activity name we have robust 
  identification, but for most of the elements this field is empty. With the same method, a \textit{content description} is sometimes associated with the element. We can also use this for strong 
  identification.        
  \item[\textbullet] \textbf{Partial identification} - If we do not have access to previous values, which happens most of the times, we can use lesser discriminative values. Textfields can 
  be set with an initial value, or a printed text can be associated to it. With no better available options we use the element dimensions to identify it. Obviously, when an element is partially
  identified, a risk of false positive is possible. 
\end{itemize} 

\noindent{}Moreover each element carries a list of actions it can trigger. Our automatic exploration consists of systematically triggering all actions of all elements for all activities. 
We do not try each combination of actions as it would not scale and be mostly redundant. To this basic general process we add targeted actions to trigger more sophisticated behaviors:

\begin{itemize} 
  \item[\textbullet] Some textfields of interest are detected like phone number, first or last name, email address, IBAN, country, city, street addresses, password or pin code. 
  These textfields are filled with consistent values accordingly. For this task, we use databases of realistic data (samples can be found in the annexes). Uncategorised textfields are filled 
  with a pseudo-random string.  
  \item[\textbullet] The order of actions done matters. For example login and password textfields must be filled before validating. In the exploration, filling textfields and 
  check-boxes takes precedence over the rest.
  \item[\textbullet] An application can register a receiver for a broadcast \textit{Android} event like the change of phone state or wifi state. It can be done statically in the application
  manifest or dynamically. Those dynamical receivers could be hidden from static analysis with obfuscation. To trigger the receivers code, we test applications with a list of broadcast 
  events that are often used by malware (a partial list is given in annexes). Moreover, real SMS and phone calls are sent to the real device own number.  
\end{itemize}

\noindent{}We finally use the \textit{Monkey} \cite{monkey} program during the analysis. It can help to trigger behaviors requiring complex inputs combination that 
\textit{Smart Monkey} could miss. At the end comes the cleaning phase. For our real device we keep a white list of regular processes and installed applications 
- regular, system and device administrator applications. Non-authorised processes are killed and applications uninstalled. Important phone configurations like wifi, 
data network and sounds are reset to a predefined value. 

\section{Discussion}

\noindent{}\textbf{Toward the standardization of the evaluation of automated testing methods for Android.}\\

Research community used different strategies for automated application testing, with different evaluation methods and different datasets. 
To promote the successful strategies for future researches on the domain, we need a standard for the experimentation. Otherwise, we cannot compare the results objectively. 
The research titled \textit{Automated Test Input Generation for Android: Are We There Yet?} \cite{testing} shows the re-evaluation on the same ground of 5 published
automated testing tools for \textit{Android}. The experimental results found is far from what have been claimed in the published papers. Moreover, according to this study \textit{Monkey}
program have the best performances above all at around 53\% \textit{average coverage of statements} on 68 selected applications. It could question the contribution of the main researches
papers to the field. However, we believe the evaluation method lacks pertinence. 

To summarize, these observations reveal several problems on the experimental results:

\begin{itemize} 
  \item[\textbullet] (1) They are currently not reproductible.  
  \item[\textbullet] (2) They cannot be compared to each other.  
  \item[\textbullet] (3) They do not hightlight the contribution of the tested tool compared to \textit{Monkey} program.
\end{itemize} 
  
To answer those problems, we propose the following rules:

\begin{itemize} 
  \item[\textbullet] (2) A \textbf{common performance measure}. We propose the \textit{average coverage of basic blocks} (this measure is described in the \textit{Experimentation} part). 
  \textit{Statement coverage} (also called \textit{line coverage}) is considered as the weakest code coverage measure by specialists in software testing. This metric should 
  not be used when another one is available. For an argued reflection about coverage metrics, we refer to the paper \textit{What is Wrong with Statement Coverage} \cite{statementcov}.  
  \item[\textbullet] (1)(2) A \textbf{common dataset} and \textbf{common tools} for instrumenting the applications. We propose the \textit{AndroCoverage Dataset} \cite{androcoverage}.
  \item[\textbullet] (1)(2) A \textbf{common configuration} - \textit{Monkey} arguments, a fixed seed for every random number generator used and application versions. These information 
  are either present in annexes of this document or on the \textit{AndroCoverage} Github web page.  
  \item[\textbullet] (3) To assess the performance of the combination of both \textit{Monkey} and the evaluated tool. Comparing separately, the performance of \textit{Monkey} and the tested tool
  does not hightlight new code paths that have been triggered by the evaluated tool. A complex method would not seem successful whereas it would have triggered complex conditions that \textit{Monkey} 
  could never find. Moreover, the \textit{Monkey} program is embeded in every \textit{Android} device (real and virtual), it interacts in a very fast pace with the application and it produces 
  good results. Then, on an operational situation, it makes sense to use it in addition to any research tool. 
\end{itemize} 

\section{Experimentation}

We use the \textit{AndroCoverage Dataset} \cite{androcoverage} for our experimentation. It contains 100 applications from \textit{F-Droid}, which is a repository of 
free and open source (\textit{FOSS}) applications. They have been manually selected with the following criteria for each application:

\begin{itemize} 
  \item[\textbullet] It does not depend on a third party library or application as an automatic tool would be unable to install it. 
  \item[\textbullet] It does not depend on root privilege. To meet the requirement of a maximum of testing tools configuration, we stick with regular privilege.
  \item[\textbullet] It does not depend on local or temporary remote data. We want the application to be usable worldwide and in the long-term. This category excludes applications 
  for a temporary event or a specific country.
\end{itemize}

\noindent{}Our goal is to use applications which show a large variety of different and steady behaviors. It is why we predict that performance on the \textit{AndroCoverage Dataset} will be 
overestimated compared to the average of real applications. This dataset is to be used to compare the performance of different automated testing tools on the same ground. 

The \textit{AndroCoverage Dataset} is supplied with tools which instruments the application, adding monitoring routines for code coverage. We used them for comparing the performance between
\textit{Monkey} and \textit{Smart Monkey}. These tools are partially founded on \textit{BBoxTester} \cite{bboxtester}, a tool for measuring the code coverage for \textit{Black Box} testing of 
\textit{Android} applications. \textit{Smart Monkey} runs \textit{Monkey} at the beginning of the analysis. For a fair trial, we tested the performance of its code coverage with and without 
\textit{Monkey}. The configuration of the \textit{Monkey} tool has been described in annexes. The results are presented on Figure \ref{fig:coverage}

\begin{figure*}[!ht]\centering
\caption{Code coverage results.}
  \scriptsize
  \begin{center}
    \begin{tabular}{ | c | c | c | c | c | }
      \hline
      \thead{\textbf{Method}} & \thead{\textbf{Classes average coverage}} & \thead{\textbf{Methods average coverage}} & \thead{\textbf{Blocks average coverage}} & \thead{\textbf{Crash rate}}\\
      \hline
      \makecell{Monkey} & 32.93\% & 35.05\% & 36.32\% & 16\%\\
      \hline
      \makecell{Smart} & 34.84\% & 36.68\% & 37.73\% & 0\%\\
      \hline
      \makecell{Smart Monkey} & 37.12\% & 41.6\% & 41.23\% & 16\%\\
      \hline
    \end{tabular}
  \end{center}
  \normalsize
\label{fig:coverage}
\end{figure*}

The \textit{Monkey} program tends to generate bugs with the instrumentation. For a significant amount of applications we are unable to get the coverage rate. 
We note that the same applications crash between \textit{Monkey} and \textit{Smart Monkey} so the crash rate has no effect on the comparison of performance between both programs. 

For the analysis of the results, we focus mainly on the \textit{average coverage of basic blocks}, as it is the most significant measure. Here are definitions of the vocabulary
used in the experimentation:

\begin{itemize} 
  \item[\textbullet] A \textit{basic block} is an uninterrupted or continuous section of instructions. It means that when the first instruction of a \textit{basic block} 
  is executed, all other instructions of the block will also be executed, only one time. A \textit{basic block} begins at the start of the program or at the 
  target of a control transfert instruction (JUMP/CALL/RETURN). It ends at the next control transfert instruction.
  
  \item[\textbullet] The \textit{basic block coverage} for an application is the number of unique basic blocks executed at runtime divided by the number of unique basic blocks 
  present in the source code.   
  
  \item[\textbullet] The \textit{average coverage of basic blocks}, is the sum of the \textit{basic block coverage} of all applications divided by the number of applications.
\end{itemize} 
 
The results shows \textit{Monkey} and \textit{Smart} (\textit{Smart Monkey} without \textit{Monkey}) do no trigger the same code paths. They have approximately the same performance but 
their combination leads to an increase of 13.52\% of \textit{average coverage of basic blocks} ($\frac{smartmonkey_{ac}}{monkey_{ac}} = 1.1352$) compared to the \textit{Monkey} program only.

\section{Limitations}

Dynamic analysis systems that allows internet communications are vulnerable to fingerprinting. Our platform is not an exception. For example \textit{Bouncer} have been the target of
remote shell attacks \cite{bouncerpwn} that enabled the fingerprinting of the system. The malware gets some information on the system and sends it to a command and control server. 
Hence, the malware author can reshape trigger conditions of the logic bomb. We accepted this risk for now. A solution halfway between shutting down all communications and no filtering at all could be to strip all outgoing information - POST request contents/GET url variables/Cookies/Metadata 
fields. This could lead to a loss of behaviors and the negative impact of such solution needs to be measured. Anyway a smart malware author will eventually find a way to leak 
remote shell outputs.

The network monitoring has limitations. First, it cannot currently handle all protocols like \textit{POP}, \textit{IMAP} and \textit{FTP} so these protocols are simply blocked.
In fact the communications are parsed, to get its content, the destination and the metadata. So this parsing needs to be changed for each protocol.  
It is an impossible task of adding one by one all protocols, so we would need to measure the protocol usage and implement the most used ones. At last, there is a countermeasure to our \textit{SSL/TLS} interception,
namely \textit{certificate pinning}. The requipement of the interception is the implantation of a custom root certificate in the \textit{Android} keystore of trusted certificates. An application 
can choose to discard the \textit{Android} keystore and to embed its own. Therefore when a communication, encrypted with our custom certificate, is checked, the communication is rejected. This 
technique is used in many banking applications \cite{autocitation}. In fact the point of view of the bank is: the user \textit{OS} cannot be trusted. Although we have no evidence that it happens for malware, it 
may be used by an avant-gardist malware and other would follow the trail. It is inconvenient for malware authors to buy a certificate signed by an authority of certification, as a 
payment trace could identify them. Despite of that, it is possible to get a valid certificate from \textit{Let's Encrypt} \cite{encrypt}, or to control a legtimate server via hacking and use
it as a relay for the \textit{C\&C} server. In these cases, \textit{certificate pinning} could be used for hiding communications from analysts or interception systems. 
A counter to this technique is instrumentation. By monitoring arguments of the \textit{SSL/TLS} encryption method, one can get the plaintext communications. We have done it 
manually for some banking applications \cite{autocitation} with \textit{APImonitor} \cite{apimonitor}, but doing it automatically is another issue. 
Applications that use \textit{certificate pinning} generally embed their own library for \textit{SSL/TLS} encryption, so detecting dynamically which call is the \textit{SSL/TLS} 
encryption method can be challenging.

Last, the cleaning phase of \textit{Glassbox} fits the security needed for a prototype. However, to move to an operational situation with malware that could execute 0-day root exploits, we need a real 
factory-reset of the phone. This is why we plan to integrate the open source project \textit{BareDroid} as a part of \textit{Smart Monkey}, for its factory-reset capability on real device.  

\section{Conclusion \& Future Research}

This paper contributes to the domain of dynamic analysis system for \textit{Android} in three ways. First, we presented \textit{Glassbox} a functional prototype of a platform that uses real 
devices, controls network and GSM communications to some extends and monitors Java calls, systems calls and network communication content. Second, we experimented \textit{Smart Monkey}, 
an automatic testing tool with a \textit{Grey Box} testing strategy. We showed that it enhances the application code coverage compared to the common \textit{Black Box} testing tool called 
\textit{Monkey}. Last, we presented a method of evaluation of automated testing tools to research community. This method covers the problems of reproducibility, the comparison with other 
works and of the contribution measurement of the tool. We made the dataset available on Github under the name \textit{AndroCoverage}.

The next step is to use \textit{Glassbox} on malware/benign applications and to use the features found on a machine learning algorithm. We are working on the classification of these data with
a neural network.

\section*{Acknowledgments}

This paper have been submitted to the ICISSP workshop FORmal methods for Security Engineering (ForSE 2017).

\section*{Annexes}

\subsection*{Data samples used in Smart Monkey}

\begin{lstlisting}
$> head first-names.txt 
Aaren
Aarika
Abagael
Abagail
Abbe
Abbey
Abbi
Abbie
Abby
Abbye

$> head random-iban.txt
AL94283405797977629281563659
AL60726122350056756457999447
AL23793884960503665784521815
AL91081264763546250859672884
AL11092882957032338172366593
AL40934720875931425549598788
AL13434083187544897640510833
AL04316725870884613781699580
AL79290673310301480830031517
AL97147262313758527061137496
\end{lstlisting}

\begin{lstlisting}
$> head broadcast-events.txt
android.intent.action.BOOT_COMPLETED
android.intent.action.BATTERY_CHANGED
android.net.conn.CONNECTIVITY_CHANGE
android.intent.action.USER_PRESENT
android.intent.action.ACTION_POWER_CONNECTED
android.intent.action.ACTION_POWER_DISCONNECTED
android.intent.action.INPUT_METHOD_CHANGED
android.bluetooth.device.action.ACL_CONNECTED
android.bluetooth.device.action.ACL_DISCONNECTED
android.intent.action.GTALK_CONNECTED
\end{lstlisting}

Emails are a combination of first and last name with a well known 
email provider (gmail, yahoo etc.)\\
 
\subsection*{Sample of a Java calls capture}

\begin{lstlisting}
$> logcat
[...]
void java.lang.StringBuilder.<init>
java.lang.StringBuilder java.lang.StringBuilder.append
java.lang.StringBuilder java.lang.StringBuilder.append
java.lang.String java.lang.StringBuilder.toString
void com.energysource.szj.android.Log.i
android.os.Looper android.os.Looper.getMainLooper
void android.os.Handler.<init>
android.os.Message android.os.Message.obtain
boolean android.os.Handler.sendMessageDelayed
void android.view.ViewGroup.onAttachedToWindow
void android.view.View.onWindowVisibilityChanged
void com.energysource.szj.embeded.AdView.updateRunning
void android.os.Handler.removeMessages
void java.lang.StringBuilder.<init>
java.lang.StringBuilder java.lang.StringBuilder.append
int android.view.View.getId
java.lang.StringBuilder java.lang.StringBuilder.append
java.lang.StringBuilder java.lang.StringBuilder.append
java.lang.String java.lang.StringBuilder.toString
[...]
\end{lstlisting}

\subsection*{Sample of a network capture}

\begin{lstlisting}
[...]
<header>
  <method>R0VU</method>
  <scheme>aHR0cA==</scheme>
  <host>MTE1LjE4Mi4zMC42OA==</host>
  <port>ODA=</port>
  <path>L0dldEluZm8uYXNoeD9hcHBpZD03Z
    mZjN2JlOTJmM2M0YTdmYTA4MzUxZTNkNT
    NmOThkYSZhcHB2ZXI9Mjc2JnY9MS4wLjQ
    mY2xpZW50PTImcG49Y29tLmdwLnNlYXJj
    aCZ1c2VydmVyPTIuMCZhZHR5cGU9MiZjb
    3VudHJ5PWZyJm50PTImbW5vPTIwODE1Jn
    V1aWQ9ZmZmZmZmZmYtZWIwOS05NDcwLTU
    zY2UtYmMxYjAwMDAwMDAwJm9zPTYuMC4x
    JmRuPUFPU1Arb24rSGFtbWVySGVhZCZza
    XplPTEwODAqMTc3NiZjYz00JmNtPTM4Lj
    QwJnJhbT0xODk5NTA4a2I=
  </path>
  <http_version>
    SFRUUC8xLjE=
  </http_version>
  <host>Y2ZnLmFkc21vZ28uY29t</host>
  <Connection>
    S2VlcC1BbGl2ZQ==
  </Connection>
  <User-Agent>
    QXBhY2hlLUh0dHBDbGllbnQvVU5BVkFJT
    EFCTEUgKGphdmEgMS40KQ==
  </User-Agent>
</header>
<content />
[...]
\end{lstlisting}

\noindent{}NB: all field values are encoded in base 64\\
\newpage
\subsection*{Monkey configuration}

\begin{lstlisting}
  monkey -s 0 --pct-syskeys 0 --pct-appswitch 0 --throttle 50 -p <package-name> -v 500
  
  -s 0: The seed of the random number generator is fixed to 0
  --pct-syskeys 0: No system key events are sent, such as Home, Back, Start Call, End Call, or Volume inputs.
  --pct-appswitch 0: No startActivity() are issued as calling the instrumentation activity another time breaks it. 
  --throttle 50: The delay between events is fixed to 50 milliseconds.
  500: A total of 500 events are sent. 
\end{lstlisting}

\phantomsection
\bibliographystyle{unsrt}
\bibliography{glassbox}

\begin{thebibliography}{10}

\bibitem{marketshare}
IDC.
\newblock Smartphone os market share, 2015 q2.
\newblock [Online]
  \url{http://www.idc.com/prodserv/smartphone-os-market-share.jsp}.

\bibitem{andrubis}
M.~Lindorfer, M.~Neugschwandtner, L.~Weichselbaum, Y.~Fratantonio, V.~v.~d.
  Veen, and C.~Platzer.
\newblock Andrubis -- 1,000,000 apps later: A view on current android malware
  behaviors.
\newblock In {\em 2014 Third International Workshop on Building Analysis
  Datasets and Gathering Experience Returns for Security (BADGERS)}, pages
  3--17, Sept 2014. DOI: 10.1109/BADGERS.2014.7.

\bibitem{androcoverage}
Androcoverage dataset.
\newblock [Online] \url{https://github.com/androcoverage/androcoverage}.

\bibitem{bouncer}
Hiroshi Lockheimer.
\newblock Android and security.
\newblock [Online]
  \url{http://googlemobile.blogspot.fr/2012/02/android-and-security.html}.

\bibitem{evasionsample}
FireEye.
\newblock Android.hehe: Malware now disconnects phone calls.
\newblock [Online]
  \url{https://www.fireeye.com/blog/threat-research/2014/01/android-hehe-malware-now-disconnects-phone-calls.html}.

\bibitem{morpheus}
Yiming Jing, Ziming Zhao, Gail-Joon Ahn, and Hongxin Hu.
\newblock Morpheus: automatically generating heuristics to detect android
  emulators.
\newblock In {\em Proceedings of the 30th Annual Computer Security Applications
  Conference}, pages 216--225. ACM, 2014. DOI: 10.1145/2664243.2664250.

\bibitem{aasandbox}
T.~Bläsing, L.~Batyuk, A.~D. Schmidt, S.~A. Camtepe, and S.~Albayrak.
\newblock An android application sandbox system for suspicious software
  detection.
\newblock In {\em Malicious and Unwanted Software (MALWARE), 2010 5th
  International Conference on}, pages 55--62, Oct 2010. DOI:
  10.1109/MALWARE.2010.5665792.

\bibitem{crowdroid}
Iker Burguera, Urko Zurutuza, and Simin Nadjm-Tehrani.
\newblock Crowdroid: behavior-based malware detection system for android.
\newblock In {\em Proceedings of the 1st ACM workshop on Security and privacy
  in smartphones and mobile devices}, pages 15--26. ACM, 2011. DOI:
  10.1145/2046614.2046619.

\bibitem{smartdroid}
Cong Zheng, Shixiong Zhu, Shuaifu Dai, Guofei Gu, Xiaorui Gong, Xinhui Han, and
  Wei Zou.
\newblock Smartdroid: an automatic system for revealing ui-based trigger
  conditions in android applications.
\newblock In {\em Proceedings of the second ACM workshop on Security and
  privacy in smartphones and mobile devices}, pages 93--104. ACM, 2012.

\bibitem{droidscope}
Lok~Kwong Yan and Heng Yin.
\newblock Droidscope: seamlessly reconstructing the os and dalvik semantic
  views for dynamic android malware analysis.
\newblock In {\em Presented as part of the 21st USENIX Security Symposium
  (USENIX Security 12)}, pages 569--584, 2012.

\bibitem{appsplayground}
Vaibhav Rastogi, Yan Chen, and William Enck.
\newblock Appsplayground: automatic security analysis of smartphone
  applications.
\newblock In {\em Proceedings of the third ACM conference on Data and
  application security and privacy}, pages 209--220. ACM, 2013. DOI:
  10.1145/2435349.2435379.

\bibitem{appaudit}
Mingyuan Xia, Lu~Gong, Yuanhao Lyu, Zhengwei Qi, and Xue Liu.
\newblock Effective real-time android application auditing.
\newblock In {\em Security and Privacy (SP), 2015 IEEE Symposium on}, pages
  899--914. IEEE, 2015.

\bibitem{bresiliens}
Vitor~Monte Afonso, Matheus~Favero de~Amorim, Andr{\'e} Ricardo~Abed
  Gr{\'e}gio, Glauco~Barroso Junquera, and Paulo~L{\'\i}cio de~Geus.
\newblock Identifying android malware using dynamically obtained features.
\newblock {\em Journal of Computer Virology and Hacking Techniques},
  11(1):9--17, 2015.

\bibitem{copperdroid}
Kimberly Tam, Salahuddin~J Khan, Aristide Fattori, and Lorenzo Cavallaro.
\newblock Copperdroid: Automatic reconstruction of android malware behaviors.
\newblock In {\em NDSS}, 2015.

\bibitem{hadm}
Lifan Xu, Dongping Zhang, Nuwan Jayasena, and John Cavazos.
\newblock Hadm: Hybrid analysis for detection of malware.
\newblock
  \url{https://www.eecis.udel.edu/~lxu/resources/HADM:\%20Hybrid\%20Analysis\%20for\%20Detection\%20of\%20Malware.pdf}.

\bibitem{maline}
Marko Dimja{\v{s}}evi{\'c}, Simone Atzeni, Ivo Ugrina, and Zvonimir Rakamaric.
\newblock Evaluation of android malware detection based on system calls.
\newblock In {\em Proceedings of the 2016 ACM on International Workshop on
  Security And Privacy Analytics}, pages 1--8. ACM, 2016. DOI:
  10.1145/2875475.2875487.

\bibitem{intellidroid}
Michelle~Y Wong and David Lie.
\newblock Intellidroid: A targeted input generator for the dynamic analysis of
  android malware.
\newblock 2016.

\bibitem{apimonitor}
Droidbox - apimonitor.wiki.
\newblock [Online]
  \url{https://code.google.com/archive/p/droidbox/wikis/APIMonitor.wiki}.

\bibitem{smali}
Github - smali readme.
\newblock [Online] \url{https://github.com/JesusFreke/smali}.

\bibitem{autocitation}
Eric Filiol and Paul Irolla.
\newblock (in)security of mobile banking... and of other mobile apps.
\newblock Black Hat Asia 2015.

\bibitem{monkey}
Google.
\newblock Ui/application exerciser monkey.
\newblock [Online] \url{https://developer.android.com/studio/test/monkey.html}.

\bibitem{monkeyrunner}
Google.
\newblock Monkeyrunner.
\newblock [Online]
  \url{https://developer.android.com/studio/test/monkeyrunner/index.html}.

\bibitem{baredroid}
Simone Mutti, Yanick Fratantonio, Antonio Bianchi, Luca Invernizzi, Jacopo
  Corbetta, Dhilung Kirat, Christopher Kruegel, and Giovanni Vigna.
\newblock Baredroid: Large-scale analysis of android apps on real devices.
\newblock In {\em Proceedings of the 31st Annual Computer Security Applications
  Conference}, pages 71--80. ACM, 2015. DOI: 10.1145/2818000.2818036.

\bibitem{clamav}
Tomasz Kojm.
\newblock Clamav, 2004.

\bibitem{wot}
WoT.
\newblock Web of trust api.
\newblock [Online] \url{https://www.mywot.com/wiki/API}.

\bibitem{inetsim}
Thomas Hungenberg and Matthias Eckert.
\newblock Inetsim: Internet services simulation suite, 2013.

\bibitem{aosp}
Google.
\newblock Welcome to the android open source project!
\newblock [Online] \url{https://source.android.com/}.

\bibitem{art}
Google.
\newblock Art and dalvik.
\newblock [Online]
  \url{https://source.android.com/devices/tech/dalvik/index.html}.

\bibitem{oat}
Paul Sabanal.
\newblock Hiding behind art.
\newblock 2015.

\bibitem{binder}
Thorsten Schreiber.
\newblock Android binder - android interprocess communication.
\newblock 2011.

\bibitem{tls}
T.~Dierks and C.~Allen.
\newblock The tls protocol version 1.0, 1999.
\newblock [Online] \url{http://www.ietf.org/rfc/rfc2246.txt}.

\bibitem{uiautomator}
Google.
\newblock Uiautomator.
\newblock [Online]
  \url{https://stuff.mit.edu/afs/sipb/project/android/docs/tools/help/uiautomator/index.html}.

\bibitem{testing}
Shauvik~Roy Choudhary, Alessandra Gorla, and Alessandro Orso.
\newblock Automated test input generation for android: Are we there yet?(e).
\newblock In {\em Automated Software Engineering (ASE), 2015 30th IEEE/ACM
  International Conference on}, pages 429--440. IEEE, 2015.

\bibitem{statementcov}
By~Steve Cornett.
\newblock What is wrong with statement coverage.
\newblock [Online] \url{http://www.bullseye.com/statementCoverage.html}.

\bibitem{bboxtester}
Yury Zhauniarovich, Anton Philippov, Olga Gadyatskaya, Bruno Crispo, and Fabio
  Massacci.
\newblock Towards black box testing of android apps.
\newblock In {\em 2015 Tenth International Conference on Availability,
  Reliability and Security (ARES)}, pages 501--510, August 2015.

\bibitem{bouncerpwn}
Nicholas~J. Percoco.
\newblock Adventures in bouncerland.
\newblock 2012.

\bibitem{encrypt}
Let's encrypt.
\newblock [Online] \url{https://letsencrypt.org/}.

\end{thebibliography}

\end{document}